\documentclass[9pt,twocolumn]{revtex4}
\usepackage{amssymb,latexsym,amsmath,graphicx}

\DeclareMathOperator{\diag}{diag}
\newcommand{\dedc}[1]{\frac{\partial E}{\partial c_{#1}}}
\newcommand{\mb}[1]{\mathbf{#1}}
\newcommand{\rx}[0]{r_{\text{exp}}}

\begin{document}
\title{ Electronic Quantum Monte Carlo Calculations of Atomic Forces,
        Vibrations, and Anharmonicities }
\author{
	Myung Won Lee$^{\text{a)}}$,
	Massimo Mella$^{\text{b)}}$,
	and Andrew~M. Rappe$^{\text{a)}}$
}
\affiliation{
        $^{\textup{a)}}$The Makineni Theoretical Laboratories,
	Department of Chemistry, University of Pennsylvania,
	Philadelphia, PA 19104-6323, USA\\
	$^{\textup{b)}}$School of Chemistry, Cardiff University,
	P.O. Box 912, Cardiff CF10 3TB, United Kingdom
}
\date{April 6, 2005}

\begin{abstract}

Atomic forces are calculated for first-row monohydrides and carbon
monoxide within electronic quantum Monte Carlo (QMC).  Accurate and
efficient forces are achieved by using an improved method for moving
variational parameters in variational QMC.  Newton's method with
singular value decomposition (SVD) is combined with steepest descent
(SD) updates along directions rejected by the SVD, after initial SD
steps.  Dissociation energies in variational and diffusion QMC agree
well with experiment.  The atomic forces agree quantitatively with
potential energy surfaces, demonstrating the accuracy of this force
procedure.  The harmonic vibrational frequencies and anharmonicity
constants, derived from the QMC energies and atomic forces, also agree
well with experimental values.

\end{abstract}
\maketitle
\startpage{1}

\section{Introduction}

Quantum Monte Carlo (QMC) is an effective method
for solving the time-independent Schr\"{o}dinger equation,
and has become quite successful
in computing ground-state total energies.
The QMC method gives energies of atoms, molecules, and solids that are
comparably accurate or more accurate than traditional techniques
such as density functional theory (DFT),
multiconfiguration self-consistent field (MCSCF), or coupled cluster methods.
Although the situation for the calculation of properties
other than energies has been less favorable,
the accurate QMC calculation of atomic forces has been enabled
through the recent developments made in this area
by Assaraf and Caffarel \cite{Assaraf00p4028,Assaraf03p10536},
Filippi and Umrigar \cite{Filippi00pR16291},
Casalegno, Mella, and Rappe \cite{Casalegno03p7193},
Chiesa, Ceperley, and Zhang \cite{Chiesa05p036404}, and others.

In this paper, we extend our atomic force methodology
to all the first-row monohydrides and carbon monoxide.
In order to acquire energies and forces efficiently for these systems,
we also describe an improved algorithm for optimizing
variational Monte Carlo (VMC) wave functions.
As in our previous paper \cite{Lin00p2650},
the first and second derivatives of the variational energy
are analytically computed, and used
to perform Newton's method parameter updates with SVD.
We now propose augmenting this approach
by using the steepest descent (SD) method
in the subspace neglected by the Newton's method with SVD.
In the initial stage of parameter update,
Newton's method might give poor result
since the second derivatives include larger noise
when the parameters are far from the optimum.
So we take two SD steps before starting Newton's method.
The improved algorithm was applied to the calculation
of the ground-state energies and forces
of the first-row monohydrides and carbon monoxide.
In general, the direct application of the variational principle
yields significantly lower energy than variance minimization methods,
so minimizing the energy is advantageous.
The wave functions optimized in VMC were used
as a guiding function to compute more accurate energies and
forces in diffusion Monte Carlo (DMC).

In this paper, total energies, dissociation energies, forces, harmonic
vibrational frequencies, and anharmonicity constants are reported for
all first-row monohydrides from LiH to HF, as well as for CO.  In all
cases, the computed results agree well with experiment.  The
dissociation energies in VMC are significantly improved with respect
to a previous VMC study of the hydrides.

\section{Theoretical Background and Computational Details}

The variational parameters used in VMC
will be denoted as $c_1$, $c_2$, $\ldots$, $c_n$, here.
The VMC energy expectation value, $E_T$,
is a function of these variational parameters,
and the parameter set that minimizes $E_T$ is sought.

%
%
The SD method is useful
in the initial stages of parameter optimization in VMC,
due to the large error bars of Hessian matrix components.
One arbitrary constant is necessary to implement the SD method.
We used the following two-step scheme to find a good SD constant.

Let $\mb{Q}_0$ and $\mb{Q}$ be the vectors
composed of variational parameters before and after update, respectively:
\begin{align}
   \mb{Q}_0 &=
   \begin{pmatrix} c_{1,0} & c_{2,0} & \cdots & c_{n,0} \end{pmatrix} ^T, \\
   \mb{Q} &=
   \begin{pmatrix} c_1 & c_2 & \cdots & c_n \end{pmatrix} ^T.
\end{align}
And let $\mb{g}$ be the gradient vector of energy
with respect to the variational parameters:
\begin{equation}
   \mb{g} = 
   \begin{pmatrix} g_1 & g_2 & \cdots & g_n \end{pmatrix} ^T =
   \begin{pmatrix} \dedc{1} & \dedc{2} & \cdots & \dedc{n} \end{pmatrix} ^T.
\end{equation}
In the first update, a value, $a^{(0)}$, is chosen
as a SD constant, which is small enough
not to exhaust the downhill direction,
\begin{equation}
   \mb{Q} = \mb{Q}_0 - a^{(0)} \mb{g}(\mb{Q}_0).
\end{equation}
After the first update, VMC simulation is performed again
to get the gradient at the new parameter set, $\mb{Q}$.
If we consider only the $i$th component, the best value
for the new SD constant, $a_i^{(1)}$,
will make the gradient component, $g_i$, zero
in the next simulation, and will be given by
\begin{equation}
   a_i^{(1)} = \frac{a^{(0)}}{1 - g_i (\mb{Q}) / g_i (\mb{Q}_0) }.
\end{equation}
Although $a_i^{(1)}$ values are different from component to component,
it is usually the case that they are quite similar.
So the averaged value was used for the next update:
\begin{equation}
   a^{(1)} = \frac{1}{n} \sum_{i=1}^n a_i^{(1)} .
\end{equation}
These two steps of parameter updates using SD
reduce the energy enough to greatly reduce the error bars,
enabling the Newton's method.

%
%
If we let $\mb{H}$ represent the Hessian matrix,
the parameters can be updated according to Newton's method,
\begin{equation}\label{Eq:Newton}
   \mb{Q} = \mb{Q}_0 - \mb{H}^{-1}(\mb{Q}_0) \mb{g}(\mb{Q}_0) .
\end{equation}
Since $\mb{H}(\mb{Q}_0)$ and $\mb{g}(\mb{Q}_0)$ are calculated
in the VMC simulation, we must invert $\mb{H}(\mb{Q}_0)$
for the Newton's method update of the parameters.

%
%
It is well-known that any matrix, e.g., $\mb{H}$, can be expressed as
\begin{equation}
   \mb{H} = \mb{U} \, [\diag (w_j)] \, \mb{V}^T,
\end{equation}
where $w_j \geq 0$ and $\mb{U}, \mb{V}$ are orthogonal \cite{PressNRF77}.
For a square matrix, the inverse matrix can be obtained by
\begin{equation}
   \mb{H}^{-1} = \mb{V} \, [\diag (1/w_j)] \, \mb{U}^T .
\end{equation}
Very small values of $w_j$ lead to erroneous moves along the directions
corresponding to these components due to large $1/w_j$ terms.
For that reason, if $w_j$ is less than a certain
threshold value, $1/w_j$ is set to $0$ in the actual calculation (SVD).

SVD method has been tested for the inversion of Hessian matrix
and it consistently gives robust results in many cases.
However, the SVD method, by zeroing out small $w_j$ values,
is equivalent to abandoning the corresponding search directions,
the use of which may give better result.
So we propose a modified algorithm in which the SD method
is added for components discarded in the SVD method.

%
%
If we let $\mb{U}$ and $\mb{V}$ be equal to the square matrix
whose column vectors are the normalized eigenvectors of $\mb{H}$,
$\{w_j\}$ will be the eigenvalues of $\mb{H}$.
For values of $w_j$ that are smaller than the threshold,
$1/w_j$ can be replaced by a constant, $a$, instead of zero,
which is equivalent to the SD method
along the corresponding directions.
This method makes it possible to use the information for all directions,
some of which are discarded in SVD method,
and it can be beneficial in cases where
some eigenvalues of the Hessian matrix become close to zero,
due to the noise inherent in QMC.
In case of SVD algorithm, $w_j$ is always nonnegative,
which corresponds to the absolute value of eigenvalue of $\mb{H}$.
If any eigenvalue is negative and its absolute value is larger
than the threshold, there is a problem
that the direction corresponding to this is not discarded,
even though this does not happen so frequently.
This small problem of negative eigenvalues can be handled
by using the modified method with the same positive threshold
and zero steepest descent constant,
and we used this modified method in the actual implementation.

%
%
To construct the trial wave functions used in VMC,
the following method was used.
First, a contracted Gaussian-type function (CGTF)
was fitted to each Slater-type orbital (STO).
Ten primitive Gaussians were used for 1$s$,
eight for 2$s$ or 2$p$, and six for 3$s$, 3$p$, or 3$d$ type STOs.
The orbital exponents of STOs in the works
of Cade and Huo \cite{Cade67p614, Cade75p1} were adopted
(excluding the $f$-type orbitals).
In case of the first-row monohydrides,
each first-row atom has 29 STOs centered on it
(1$s$, 1$s'$, 2$s$, 2$s'$, 3$s$, three 2$p$'s, three 2$p'$'s,
three 2$p''$'s, three 3$p$'s, six 3$d$'s, and six 3$d'$'s for Li,
and 1$s$, 1$s'$, 2$s$, 2$s'$, 3$s$, three 2$p$'s, three 2$p'$'s,
three 2$p''$'s, three 2$p'''$'s, six 3$d$'s, and six 3$d'$'s
for other first-row atoms) and hydrogen atom has 6 STOs centered on it
(1$s$, 1$s'$, 2$s$, and three 2$p$'s) as a basis set.
In case of carbon monoxide, each atom has 19 STOs centered on it
(1$s$, 1$s'$, 2$s$, 3$s$, three 2$p$'s, three 2$p'$'s,
three 2$p''$'s, and six 3$d$'s) as a basis set.

Each molecular orbital (MO)
was expressed as a linear combination
of STOs, the coefficients of which were obtained
using the Hartree-Fock method in Gaussian 98 (G98) \cite{Gaussian98}.
For the open shell molecules, restricted open shell Hartree-Fock (ROHF)
wave functions were used.
The MOs from G98 were used to construct the Slater determinants
for $\alpha$ and $\beta$ electrons.
While multideterminant trial wave function gives
improved results for some systems, it was reported that
the use of single determinant trial wave function
gave good results in the calculations of the first-row hydrides
\cite{Luchow96p7573,Morosi99p6755}.
Since the use of multideterminant trial wave function
is much more time-consuming, we used only single determinant
in the calculation here.
The product of two determinants was multiplied
by a positive correlation factor
to form a trial wave function \cite{Boys69p43,Schmidt90p4172}:
\begin{equation}
   \Psi_T = D^{\uparrow} D^{\downarrow} \exp(\sum_{a,i<j}U_{aij}),
\end{equation}
where
\begin{equation}
   U_{aij} = \sum_k^{N_a} c_{ka} ( \bar{r}_{ai}^{l_{ka}} \bar{r}_{aj}^{m_{ka}}
   + \bar{r}_{aj}^{l_{ka}} \bar{r}_{ai}^{m_{ka}} ) \bar{r}_{ij}^{n_{ka}}.
\end{equation}
In this equation, $a$ and $i,j$ refer to the nuclei and the electrons,
respectively, and $\bar{r}$ is defined by $\bar{r}=br/(1+br)$.
$c_{ka}$'s are variational Jastrow parameters.
We used $b=1 \: a_0^{-1}$ and included 30 terms for diatomic molecules,
namely, 4 electron-electron, 6 electron-nucleus,
and 20 electron-electron-nucleus terms.
In case of atoms, we used 17 parameters composed of 
4 electron-electron, 3 electron-nucleus,
and 10 electron-electron-nucleus terms,
to be consistent with the calculation of diatomic molecules.

%
%
Five different bond distances around the experimental bond length
were used for calculation, namely 90\%, 95\%, 100\%, 105\% and 110\%
of the experimental bond length, $\rx$.
2000 walkers were used for all the calculations in this paper.
In updating Jastrow parameters,
average over 100 blocks was made typically,
where each block was the average over 100 steps.
To accelerate the sampling, a Fokker-Planck type equation
was used \cite{HammondAQC}.

After a short initial simulation without Jastrow factor,
the Hartree-Fock wave function was multiplied
by the Jastrow factor with all parameters set to zero.
The gradient and Hessian of energy with respect to
the Jastrow parameters were computed
in the VMC simulation after this step.
Using the gradient and Hessian information,
a new Jastrow parameter set is calculated,
and a new VMC simulation is performed with this updated parameter set.
This process was iterated until the energy converged.
Fully optimized parameters were obtained by 10-15 iterations.
One iteration took about 30 minutes for LiH and about 90 minutes for HF
when a single 2.8 GHz Intel$^{\text{\textregistered}}$
Xeon$^{\text{\texttrademark}}$ Processor was used.

After optimizing the trial wave function using VMC,
a fixed-node DMC calculation was performed using importance sampling,
as proposed by Reynolds, Ceperley, Alder, and Lester \cite{Reynolds82p5593}.
The DMC time step was 0.005 a.u. for the first-row hydrides
and 0.0005-0.001 a.u. for carbon monoxide.
A similar DMC method was used by L\"{u}chow and Anderson
\cite{Luchow96p4636,Luchow96p7573}
in their calculation of first-row hydrides.

%
%
Force calculations were performed in both VMC and DMC.
We followed the method described previously \cite{Casalegno03p7193}.
If the wave function were exact, the exact force would be given
by the Hellmann-Feynman theorem (HFT).
Since the trial wave function, $\Psi_T$, is not exact,
terms that cancel in case of exact wave functions should be considered,
in addition to the HFT expression.
Retaining terms involving wave function derivatives
gives the total atomic force on atom $a$ in direction $q$:
\begin{equation}
   F_{qa} = F_{qa}^{\text{HFT}} + F_{qa}^{\text{Pulay}} + F_{qa}^{c},
\end{equation}
where
\begin{equation}
   F_{qa}^{\text{HFT}}
   = - \frac{\langle \Psi_T | \frac{\partial \Hat{H}}{\partial R_{qa}}
     |\Psi_T \rangle}
   {\langle \Psi_T | \Psi_T \rangle} ,
\end{equation}
\begin{equation}
   F_{qa}^{\text{Pulay}}
   = - 2 \frac{\langle \frac{\partial \Psi_T}{\partial R_{qa}} |
   \Hat{H} | \Psi_T \rangle} {\langle \Psi_T | \Psi_T \rangle}
   + 2 \langle E \rangle_{\text{VMC}}
   \frac{\langle \frac{\partial \Psi_T}{\partial R_{qa}} | \Psi_T \rangle}
   {\langle \Psi_T | \Psi_T \rangle} ,
\end{equation}
and
\begin{equation}
   F_{qa}^{c}
   = - \sum_k \frac{\partial c_k}{\partial R_{qa}}
   \frac{\partial \langle E \rangle_{\text{VMC}}}{\partial c_k}.
\end{equation}
These expressions apply for VMC,
and similar equations are used for DMC simulations \cite{Casalegno03p7193}.
$F_{qa}^{\text{Pulay}}$ incorporates the explicit dependence
of the wave function on the nuclear coordinates
(Pulay's correction \cite{Pulay69p197}),
and can be easily calculated through VMC or DMC simulations.
$F_{qa}^{c}$ depends implicitly
on the nuclear coordinates through the variational parameters.
However, since an energy-minimized wave function is used,
i.e., $\partial \langle E \rangle_{\text{VMC}} / \partial c_k = 0$,
this force term makes zero contribution.
In the calculation of the Hellmann-Feynman theorem force, $F_{qa}^{\text{HFT}}$,
the renormalized estimator proposed by Assaraf and Caffarel
\cite{Assaraf00p4028}
was used to reduce the variance of the force calculation.
The expectation value of this estimator, $F_{qa}^{\text{AC}}$,
is the same as $F_{qa}^{\text{HFT}}$,
but the variance of the former is much smaller.
In our force calculation,
$F_{qa}^{\text{AC}}$ + $F_{qa}^{\text{Pulay}}$
was computed by averaging over the walkers.

\section{Results and Discussion}

\begin{figure} [b]  
   \centerline{\includegraphics[width=2.75in]{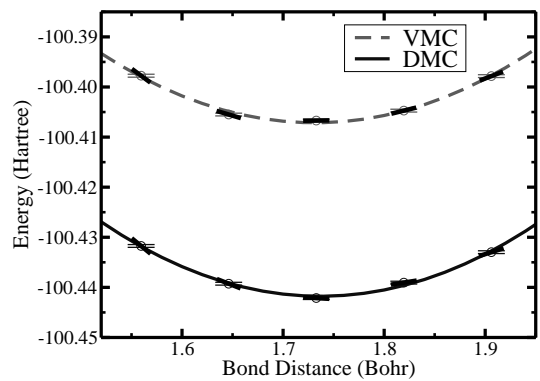}}
   \caption{ Energy and force calculation of HF with VMC and DMC.
             Two thin horizontal lines at each data point
             show the energy error bar.
             The slope of the thick lines show the force
             at each data point.}
   \label{Fi:HFplot}
\end{figure}

%
%
The energies of first-row monohydrides and carbon monoxide
at various bond distances were calculated.
The plot of energy versus bond distance for hydrogen fluoride (HF)
is shown in Figure~\ref{Fi:HFplot}.
In obtaining each point,
1000 blocks, each of which was composed of 100 steps,
were used with optimized Jastrow parameters.
The plots for other molecules are similar to that for HF.
The energies obtained from VMC are a few tenths of a Hartree lower
than the Hartree-Fock energies obtained from G98,
so the Hartree-Fock results are not shown in the figure.
It can be seen from Table~\ref{Tb:Morsefit}
that the DMC energy is significantly lower than the VMC energy
and is close to the experimental value.

\begin{table}  
\caption{$E_0$, $r_e$, $\omega_e$, and $\omega_e x_e$ for LiH - HF and CO
         obtained from VMC and DMC calculation and experimental data.}
\begin{center}
   \begin{tabular}{llcrcllll}
      \hline \hline
      & & & $E_0$ (Ha) && $r_e$ (Bohr) & $\omega_e$ (cm$^{-1}$)
          & $\omega_e x_e$ (cm$^{-1}$) \\ \hline
      LiH \qquad
      & VMC  &&   -8.063 && 3.038(1) & 1402(4) & 25.7(1) \\
      & DMC  &&   -8.070 && 3.020(1) & 1417(4) & 24.8(1) \\
      & Exp  &&   -8.070 && 3.015    & 1406    & 23.2 \\
      \hline
      BeH \qquad
      & VMC  &&  -15.235 && 2.519(1) & 2141(4) & 56.6(2) \\
      & DMC  &&  -15.246 && 2.515(1) & 2134(4) & 58.5(2) \\
      & Exp  &&  -15.248 && 2.537    & 2061    & 36.3 \\
      \hline
      BH \qquad
      & VMC  &&  -25.254 && 2.370(1) & 2332(5) & 47.0(2) \\
      & DMC  &&  -25.275 && 2.386(1) & 2369(5) & 47.3(2) \\
      & Exp  &&  -25.289 && 2.329    & 2367    & 49.4 \\
      \hline
      CH \qquad
      & VMC  &&  -38.438 && 2.097(1) & 2961(6) & 77.2(3) \\
      & DMC  &&  -38.463 && 2.112(1) & 2898(6) & 71.8(3) \\
      & Exp  &&  -38.490 && 2.116    & 2858    & 63.0 \\
      \hline
      NH \qquad
      & VMC  &&  -55.178 && 1.941(1) & 3415(7) & 104.3(4) \\
      & DMC  &&  -55.206 && 1.962(1) & 3253(7) & 92.0(4) \\
      & Exp  &&  -55.247 && 1.958    & 3282    & 78.4 \\
      \hline
      OH \qquad
      & VMC  &&  -75.687 && 1.820(1) & 3854(7) & 101.2(4) \\
      & DMC  &&  -75.720 && 1.843(1) & 3690(7) & 91.4(4) \\
      & Exp  &&  -75.778 && 1.832    & 3738    & 84.9 \\
      \hline
      HF \qquad
      & VMC  &&  -100.407 && 1.729(1) & 4206(9) & 89.9(4) \\
      & DMC  &&  -100.442 && 1.755(1) & 4040(9) & 82.4(4) \\
      & Exp  &&  -100.531 && 1.733    & 4138    & 89.9 \\
      \hline
      CO \qquad
      & VMC  &&  -113.176 && 2.095(1) & 2539(16) & 21.1(3) \\
      & DMC  &&  -113.286 && 2.116(2) & 2251(26) & 14.2(3) \\
      & Exp  &&  -113.377 && 2.132    & 2170     & 13.3 \\
      \hline \hline
   \end{tabular}
\end{center}
\label{Tb:Morsefit}
\end{table}

%
%
The bond dissociation energies, $D_e$, were calculated by
taking the differences between QMC energies of diatomic molecules
in Table~\ref{Tb:Morsefit} and QMC energies of atoms.
To be consistent in the number of Jastrow parameters,
we did the calculation of atoms with 17 parameters.
The VMC energies of atoms with 17 parameters falls
between those with 9 parameters and those with 42 parameters
reported in Ref. \cite{Lin00p2650}.
The dissociation energies are summarized
in Table~\ref{Tb:dissocE}, together with the results
given in the work by L\"{u}chow and Anderson \cite{Luchow96p7573}.
Our VMC dissociation energies are much closer to experimental values
than those given by L\"{u}chow and Anderson,
while our DMC results are quite similar to theirs.
The improvement in our VMC result may be attributed to the effectiveness
of energy minimization method relative to the variance minimization method
used for VMC calculations in Ref. \cite{Luchow96p7573},
while part of the improvement is also due to the larger
number of Jastrow parameters in our calculation.

\begin{table}  
\caption{Dissociation energies $D_e$ in kcal/mol for LiH - HF and CO
         obtained from our QMC calculation and from literature.}
\begin{center}
   \begin{tabular}{cccccccc}
      \hline \hline \quad
    && \: VMC\footnotemark[1] \: &
       \: DMC\footnotemark[1] \: &
       \: VMC\footnotemark[2] \: &
       \: DMC\footnotemark[2] \: &
       \: Exp\footnotemark[2] \: \\ \hline
LiH &&	 54.7  &  57.8  &   45.7 &   57.8  &  58.0 \\
BeH &&	 57.9  &  55.7  &   49.4 &   52.1  &  49.8 \\
BH  &&	 82.7  &  84.7  &   63   &   84.8  &  84.1 \\
CH  &&	 81.1  &  83.5  &   81   &   83.9  &  83.9 \\
NH  &&	 80.2  &  82.3  &   77   &   81.4  &  80.5-84.7 \\
OH  &&	105.1  & 106.4  &   86   &  106.4  &  106.6 \\
HF  &&	140.4  & 141.4  &  130   &  141.3  &  141.5 \\
CO  &&	218.1  & 254.9  &   -    &    -    &  258.7 \\
      \hline \hline
\footnotetext[1]{
         Differences between QMC energies of molecules in Table~\ref{Tb:Morsefit}
         and QMC energies of atoms calculated with 17 parameters. \\
}
\footnotetext[2]{
         From Ref. \cite{Luchow96p7573} for first-row hydrides.
}
   \end{tabular}
\end{center}
\label{Tb:dissocE}
\end{table}

%
%
Energies calculated by DMC are quite close to the experimental values
for lighter first-row hydrides,
while slightly higher energies than experimental values are obtained
for heavier molecules.
This may be due to the approximations used in DMC calculations:
fixed node approximation, neglect of the relativistic effect,
and the error related with finite time step.
To estimate the finite time step error,
DMC calculations at $\rx$ with several different time step values
ranging from 0.0001 to 0.005 a.u. were carried out for first-row hydrides.
All calculated energies agreed within 2-3 mHartree.

%
%
In the VMC calculation of HF, the Jastrow parameter set
at $\rx$ was optimized first,
and after the optimization at this distance,
the bond distance was changed, and the MO coefficients
corresponding to this bond distance were introduced.
Then, the Jastrow parameters were reoptimized at this new bond distance.
This method makes it possible to reduce the CPU time
for the calculation at other bond distances
once the parameter set is optimized at one bond distance.
This approach is effective because the Jastrow parameter sets
at different bond distances can be quite similar, as measured
by the cosine similarity \cite{SaltonMIR}
between Jastrow parameter sets,
\begin{equation}
   \cos \theta =
   \frac{\mb{Q}_m \cdot \mb{Q}_n}
   {\sqrt{\mb{Q}_m \cdot \mb{Q}_m} \sqrt{\mb{Q}_n \cdot \mb{Q}_n}},
\end{equation}
which is close to unity if two vectors are similar.
This is certainly the case for
Jastrow parameter sets of HF at various bond distances,
as shown in Table~\ref{Tb:cosine}.
This approach seems to be useful for the molecular dynamics (MD) simulation
coupled with QMC, proposed by Grossman and Mitas \cite{Grossman05p056403}.
On the other hand, in case of CH, NH, or OH,
it was problematic to apply this method
and we had to optimize the parameters from the beginning for all bond distances.
The cosine similarity values in case of CH are shown in Table~\ref{Tb:cosine},
when the parameters are optimized separately from scratch
for all bond distances.
If the parameters of HF at each bond distance are optimized from scratch,
the cosine similarity values are around 0.9 for parameter sets optimized at
different bond distances, and similar energies can be obtained
with different sets of parameters.

\begin{table}  
\caption{The cosine similarity values between Jastrow parameter sets
         obtained from VMC calculations of HF and CH.}
\begin{center}
   \begin{tabular}{c|ccccccc}
      \hline \hline
      $\cos \theta$ (HF) && $0.90\,\rx$ & $0.95\,\rx$ &
      $1.00\,\rx$ & $1.05\,\rx$ & $1.10\,\rx$ \\ \hline
      $0.90\,\rx$   && 1.000  &        &        &        &        \\
      $0.95\,\rx$   && 0.998  & 1.000  &        &        &        \\
      $1.00\,\rx$   && 0.997  & 1.000  & 1.000  &        &        \\
      $1.05\,\rx$   && 0.998  & 0.997  & 0.997  & 1.000  &        \\
      $1.10\,\rx$   && 0.997  & 0.999  & 0.999  & 0.997  & 1.000  \\
      \hline
      \hline
      $\cos \theta$ (CH) && $0.90\,\rx$ & $0.95\,\rx$ &
      $1.00\,\rx$ & $1.05\,\rx$ & $1.10\,\rx$ \\ \hline
      $0.90\,\rx$   && 1.000  &        &        &        &        \\
      $0.95\,\rx$   && 0.836  & 1.000  &        &        &        \\
      $1.00\,\rx$   && 0.842  & 0.964  & 1.000  &        &        \\
      $1.05\,\rx$   && 0.658  & 0.807  & 0.701  & 1.000  &        \\
      $1.10\,\rx$   && 0.829  & 0.879  & 0.848  & 0.817  & 1.000  \\
      \hline \hline
   \end{tabular}
\end{center}
\label{Tb:cosine}
\end{table}

\begin{figure} [b]  
   \centerline{\includegraphics[width=2.75in]{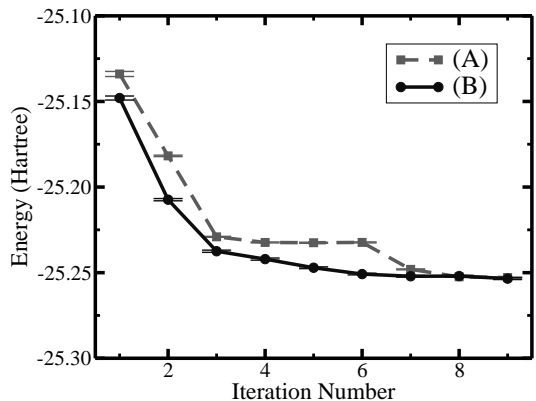}}
   \caption{ The energy of BH at $\rx$ at various stages of
             parameter optimization. (A) Newton's method for 1-6
             and Newton's method with SD for 6-9.
             (B) Initial SD for 1-3 and Newton's method for 3-9.}
   \label{Fi:BHiter}
\end{figure}

%
%
The energy of BH at $\rx$ at various stages of parameter optimization
is shown in Figure~\ref{Fi:BHiter}.
If the SD steps are used for initial stages of parameter optimization (B),
Newton's method with SVD converges to the lowest energy after several iterations.
If the initial SD steps are not used (A), Newton's method is somewhat
difficult to apply due to the large error bars of Hessian components.
In this case, it was necessary
to set the SVD threshold somewhat high
and to calculate for a long period of time.
Within this approach, using only the Newton's method with SVD
does not yield fully optimized energy.
The simultaneous application of Newton's method and SD (steps 6-9)
was very useful in this case for more thorough minimization.

%
%
Forces were computed for each monohydride and carbon monoxide
at each bond length in VMC and DMC.
The force result for HF is shown in Figure~\ref{Fi:HFplot},
where the slopes of the line segments superimposed on the energy result
correspond to the negative of the calculated forces.
The calculated forces of HF are shown in Table~\ref{Tb:forcecf},
together with the values obtained by fitting energy result
to the parabolic potential and then calculating the slopes.
The force at $0.90 \, \rx$ is larger than the magnitude of
the slope of the parabola obtained from the energy result,
while the magnitude of the force at $1.10 \, \rx$ is smaller
than the parabola tangent, which clearly shows the deviation
of the calculated forces from harmonic behavior due to anharmonicity.

\begin{table*}  
\caption{Forces obtained from the slope of parabolic potential energy fits
         and from the direct calculation for HF.}
\begin{center}
   \begin{tabular}{rrlllllll}
      \hline \hline
      & Force & \quad & \: & \; $0.90\,\rx$ \: & \; $0.95\,\rx$ \: &
         \; $1.00\,\rx$ \: & \; $1.05\,\rx$ \: & \; $1.10\,\rx$ \: \\ \hline
      & VMC & (parabola) &
         &  0.113(11) &  0.057(6) &  0.001(3) & -0.055(6) & -0.111(11) \\
      & VMC & (direct) &
         &  0.147(1)  &  0.060(1) & -0.002(1) & -0.050(1) & -0.076(2) \\ \hline
      & DMC & (parabola) &
         &  0.110(4)  &  0.056(2) &  0.002(1) & -0.051(2) & -0.105(4) \\
      & DMC & (direct) &
         &  0.168(1)  &  0.077(1) &  0.015(1) & -0.033(1) & -0.064(1) \\ \hline
      \hline
   \end{tabular}
\end{center}
\label{Tb:forcecf}
\end{table*}

%
%
The approximate shape of the anharmonic potential can be described
by the Morse potential \cite{Morse29p57},
\begin{equation}
   V(r) = D_e \left( 1 - e^{-\beta (r-r_e)} \right)^2,
\end{equation}
and this was used in the fitting of the QMC results
to calculate the properties of diatomic molecules.
The $v$th energy level of the Morse potential
with reduced mass $\mu$ is
\begin{equation}
   \frac{E}{hc} = \omega_e \left( v + \frac{1}{2} \right)
                - \omega_e x_e \left( v + \frac{1}{2} \right)^2,
\end{equation}
where the harmonic vibrational frequency is given by
$\omega_e = \beta ( 100D_e h / 2 \pi^2 c \mu )^{1/2}$
and the anharmonicity constant by
$\omega_e x_e = (100 h \beta^2 / 8 \pi^2 \mu c)$.
In this equation, $\omega_e$, $D_e$ and $\beta$ have the unit of cm$^{-1}$
and other constants are in SI units.

%
%
Since we performed QMC calculations at small number of bond distances,
it would be advantageous to use the energy and force results simultaneously
in the fitting to the Morse potential,
which can be accomplished by minimizing the following merit function:
\begin{equation}
   \chi^2 (\mb{a})
   = \sum_{i=1}^{N} \left[ \frac{E_i - E(r_i; \mb{a})} {\sigma_{E,i}} \right]^2
   + \sum_{i=1}^{N} \left[ \frac{F_i - F(r_i; \mb{a})} {\sigma_{F,i}} \right]^2.
\end{equation}
Here $\mb{a}$ is a parameter vector
whose components are $D_e$, $\beta$, and $r_e$,
and the following functional forms were used:
\begin{align}
   E(r;\mb{a}) &= D_e \left[ \left( 1-e^{-\beta(r-r_e)} \right)^2 - 1 \right]
                + \left( E_\text{A} + E_\text{B} \right), \\
   F(r;\mb{a}) &= -2 D_e \beta \left( 1-e^{-\beta(r-r_e)} \right) e^{-\beta(r-r_e)}.
\end{align}
The expression for the energy has been modified
to produce correct dissociation energy, $D_e$,
and $E_\text{A}$ and $E_\text{B}$ are VMC or DMC energies of atoms A and B.
The equilibrium bond lengths ($r_e$),
harmonic vibrational frequencies ($\omega_e$),
and anharmonicity constants ($\omega_e x_e$)
for all first-row monohydride molecules and carbon monoxide
calculated by fitting energy and force results
are summarized in Table~\ref{Tb:Morsefit},
along with the experimental data \cite{HerzbergCDM}.
The experimental energies are corrected by adding zero point energies.
Our calculations agree well with the experimental results.

%
%
Each energy or force data point has an error bar associated with it,
so we followed a simple procedure to estimate
how the calculated error bars translate
into uncertainty in other quantities
such as equilibrium bond length, harmonic vibrational frequency,
and anharmonicity constant.
A large set of synthetic data points were stochastically generated,
such that the average value at each bond length agrees with that
obtained from QMC with the standard deviation
the same as the error bar given by the QMC calculation.
By computing the averages and standard deviations of
the equilibrium bond length ($r_e$),
harmonic vibrational frequency ($\omega_e$),
and anharmonicity constant ($\omega_e x_e$)
for the synthetic data sets,
the error bars of $r_e$, $\omega_e$ and $\omega_e x_e$ could be estimated.
The error bars of the last digit thus calculated are shown in parentheses.

\section{Conclusions}

The force calculation method
combining energy minimization, Pulay's corrections,
and a renormalized Hellmann-Feynman estimator
worked well with all the first-row hydride molecules and carbon monoxide
with small extra effort.

The energy minimization method in VMC is useful,
but it requires an effective optimization scheme.
The addition of steepest descents to the initial steps
and to the subspace neglected by Newton's method with SVD
seems to be advantageous for the molecular systems we investigated.

We could calculate accurate harmonic vibrational frequencies
and anharmonicity constants of diatomic molecules
by fitting QMC results to the Morse potential,
achieving excellent agreement between QMC calculations
and experiment for these vibrational parameters.

\section{Acknowledgments}

This work was supported
by the Air Force Office of Scientific Research,
under Grant No. FA9550-04-1-0077,
and the Office of Naval Research,
under Grant No. N-000014-00-1-0372.
Computational support was provided by the National Science Foundation
CRIF Program, Grant 0131132.

\bibliography{qmc1_arx}

\end{document}